\documentclass{article}

\usepackage[final, nonatbib]{neurips_2023}

\author{%
    Ilaria Manco\thanks{Equal contribution} $^{1,2}$, Benno Weck$^{*3}$, Seungheon Doh$^4$, Minz Won$^5$,\\
     \textbf{Yixiao Zhang$^1$, Dmitry Bogdanov$^3$, Yusong Wu$^6$, Ke Chen$^7$, Philip Tovstogan$^3$}\\
    \textbf{Emmanouil Benetos$^1$, Elio Quinton$^2$, George Fazekas$^1$, Juhan Nam$^4$} \\
    $^1$QMUL, $^2$UMG, $^3$UPF, $^4$KAIST, $^5$ByteDance, $^6$Mila, $^7$UCSD
}

\usepackage[utf8]{inputenc} 
\usepackage[T1]{fontenc}    
\usepackage[colorlinks=true,citecolor=blue]{hyperref}   
\usepackage{url}            
\usepackage{booktabs}       
\usepackage{amsfonts}       
\usepackage{nicefrac}       
\usepackage{microtype}      
\usepackage{xcolor}         
\usepackage{tabularx}
\usepackage{amsmath,graphicx}
\usepackage{multirow, tabularx, booktabs, caption}
\usepackage{arydshln}
\usepackage{multirow}
\usepackage{pifont}
\usepackage{subcaption}
\usepackage[font=small]{caption}
\usepackage{enumitem}
\newcommand{\xmark}{\ding{55}}%

\makeatletter
\def\adl@drawiv#1#2#3{%
        \hskip.5\tabcolsep
        \xleaders#3{#2.5\@tempdimb #1{1}#2.5\@tempdimb}%
                #2\z@ plus1fil minus1fil\relax
        \hskip.5\tabcolsep}
\newcommand{\cdashlinelr}[1]{%
  \noalign{\vskip\aboverulesep
           \global\let\@dashdrawstore\adl@draw
           \global\let\adl@draw\adl@drawiv}
  \cdashline{#1}
  \noalign{\global\let\adl@draw\@dashdrawstore
           \vskip\belowrulesep}}
\makeatother

\title{The Song Describer Dataset: a Corpus of Audio Captions for Music-and-Language Evaluation}
\begin{document}

\maketitle

\begin{abstract}
We introduce the Song Describer dataset (SDD), a new crowdsourced corpus of high-quality audio-caption pairs, designed for the evaluation of music-and-language models. The dataset consists of 1.1k human-written natural language descriptions of 706 music recordings, all publicly accessible and released under Creative Common licenses. 
To showcase the use of our dataset, we benchmark popular models on three key music-and-language tasks (music captioning, text-to-music generation and music-language retrieval). Our experiments highlight the importance of cross-dataset evaluation and offer insights into how researchers can use SDD to gain a broader understanding of model performance.
\end{abstract}

\section{Introduction} \label{sec:introduction}
Multimodal approaches that jointly process audio and language are becoming increasingly important within music understanding and generation, giving rise to a new area of research, which we refer to as music-and-language (M\&L). Several recent works have emerged in this domain, proposing methods to automatically generate music descriptions \cite{Manco2021, doh_lp-musiccaps_2023, gabbolini_data-efficient_2022}, synthesise music from a text prompt \cite{agostinelli_musiclm_2023, huang_noise2music_2023, schneider_mousai_2023, copet_simple_2023}, search for music based on language queries \cite{doh_toward_2022, Manco2022, huang_mulan_2022}, and more \cite{liu_music_2023, leszczynski_conversational_2022}. However, evaluating M\&L models remains a challenge due to a lack of public and accessible datasets with paired audio and language, resulting in the widespread use of private data \cite{Manco2021, Manco2022, Manco2021a, huang_noise2music_2023, agostinelli_musiclm_2023, huang_mulan_2022} and inconsistent evaluation practices. To mitigate this, we release the Song Describer dataset (SDD), a new high-quality evaluation dataset of crowdsourced captions paired with openly licensed music recordings. Through our dataset, we allow for the first standardised comparison of models on realistic music-language data and expand the number and variety of datasets available to the research community. We propose SDD as an evaluation-only dataset to promote the development and use of out-of-domain data and counteract the tendency to overfit to a particular dataset \cite{bowman_what_2021, recht_imagenet_2019, vasudevan_when_2023}.

Alongside SDD, there are currently two other human-curated datasets of music-caption pairs: MusicCaps \cite{agostinelli_musiclm_2023} and YT8M-MusicTextClips 
\cite{mckee_language-guided_2023}, composed of 5.5k and 4k 10-second clips from AudioSet \cite{Gemmeke2017} and YouTube8M \cite{abu-el-haija_youtube-8m_2016} respectively. More recently, Doh et al. \cite{doh_lp-musiccaps_2023} have released LP-MusicCaps, a dataset of 2.2M music descriptions generated by prompting a large language model with a set of instructions and ground-truth tags. While synthetic data presents an opportunity to artificially scale up training data, limitations such as LLM hallucinations still pose significant challenges to their use in evaluation, where reliable data plays a crucial role.

\section{The Song Describer Dataset} \label{sec:song_describer}
SDD is a corpus of manually created descriptions for high-quality audio recordings from the MTG-Jamendo dataset~\cite{Bogdanov2019}, all available under Creative Commons licenses, and distributed as part of the dataset alongside the annotations, which we release under the CC BY-SA 4.0 license.\footnote{\url{https://doi.org/10.5281/zenodo.10072001}} Captions in SDD contain syntactically and semantically rich single-sentence descriptions covering various musical features such as instrumentation, genre, style and mood, as shown in the examples in Table \ref{tab:sd_examples}.
Notably, recordings in SDD can also be connected to additional metadata from MTG-Jamendo, enriching this with tags of several musical categories (see Figure \ref{fig:tag_distribution} in Appendix \ref{appendix:datasheet_composition}).

Compared to MusicCaps and YT8M-MusicTextClips, SDD is characterised by (1) annotations of longer track segments (95\% of the tracks are 2 minutes long); (2) openly accessible audio, persistently stored, and with explicit licenses; (3) a larger pool of annotators, composed of non-experts with diverse backgrounds and more representative of the general public (this is only known for MusicCaps); (4) a portion of recordings (25\%) with more than one caption produced by different annotators, making it particularly suitable for evaluation via automatic metrics for captioning. In contrast, audio clips are not distributed as part of MusicCaps and YT8M-MusicTextClips and can only be indirectly obtained via YouTube, with no guarantees about data persistence. Moreover, audio in MusicCaps is derived from AudioSet, which is often used for pre-training large audio models, thus posing a risk of data leakage and in-distribution bias \cite{weck_data_2023}. We summarise these and additional key differences in Table \ref{tab:mc_vs_sd}. 

\begin{table*}[t]
\centering
\caption{
An overview of the SDD compared to other music-caption datasets, MusicCaps (MC) \cite{agostinelli_musiclm_2023} and YT8M-MusicTextClips (MTC) \cite{mckee_language-guided_2023}. * denotes the validated subset.}
\footnotesize
\begin{tabular}{p{0.7cm}ccccccccc}
\toprule
\multirow{2}{*}{Dataset} & \multirow{2}{*}{Annotators} & \multicolumn{3}{c}{Text} & \multicolumn{4}{c}{Audio} \\
\cmidrule(rl){3-5} \cmidrule(rl){6-9}
 & & Caption $\#$ & Avg length & Vocab size & Audio $\#$ & Length & Source & Public \\
\midrule
MC &  10 & 5,521 & 54.8 $ \pm$ 18.7  & 6,144 & 5,521 & 10 sec & \cite{Gemmeke2017} & \xmark \\
MTC &  - & 4,169 & 16.9 $ \pm$ 4.4 & 2,599 & 4,169 & 10 sec & \cite{abu-el-haija_youtube-8m_2016} & \xmark \\
\cdashlinelr{1-9}
SDD* & 114 & 746 & 18.2 $\pm$ \phantom{0}7.6  & 1,942 & 547 & 2 min & \cite{Bogdanov2019} & \checkmark \\
SDD & 142 & 1106 & 21.7 $\pm$ 12.4  & 2,859 & 706 & 2 min & \cite{Bogdanov2019} & \checkmark \\
\bottomrule
\end{tabular}
\label{tab:mc_vs_sd}
\end{table*}

\begin{table*}[t]
\centering
\caption{The main task instruction given to the annotators, together with an example of three captions produced by different annotators describing the same track in the dataset.}
\footnotesize
\begin{tabular}{p{3.15cm}p{10cm}}
\toprule
Instructions & Example caption \\
\midrule
\multirow{7}{3.15cm}{Write one sentence that describes the track you just listened to: focus on descriptive text which conveys the overall content of this track only.} 
& \textit{A folky song with a warm, organic and cosy sound, sung by a single male voice and accompanied by a number of acoustic guitars.} \\
  \cdashlinelr{2-2}
& \textit{This is a chill folk song with a simple rythm and acoustic guitar, good for a slow morning of reading or working.} \\
  \cdashlinelr{2-2}
& \textit{A natural, singer-songwriter esque song driven by croaky male vocals, acoustic guitars, and found drum sounds; later a shaker emerges as the singer uplifts the melody with his voice.} \\
\bottomrule
\end{tabular}
 \vspace{-4mm}
\label{tab:sd_examples}
\end{table*}

\begin{figure}
    \centering
    \captionbox{Distribution of the top nine countries of residence contributors reported in the on-boarding survey. \label{fig:countries}}
[.48\textwidth]{\includegraphics[width=0.4\textwidth]{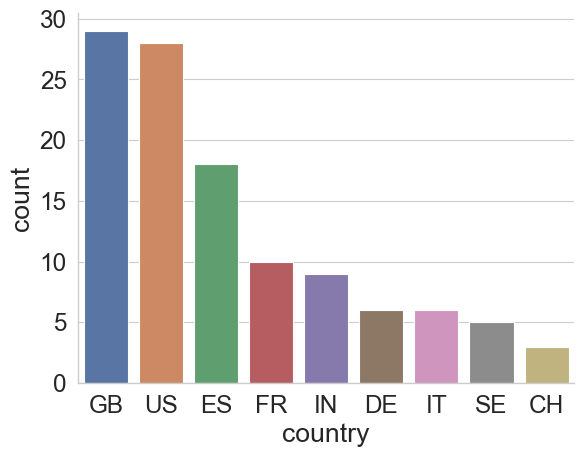}}%
    \hfill
    \captionbox{Distribution of music aspects covered by the most frequent word stems in the collected captions. \label{fig:facets}}
[.48\textwidth]{\includegraphics[width=0.45\textwidth]{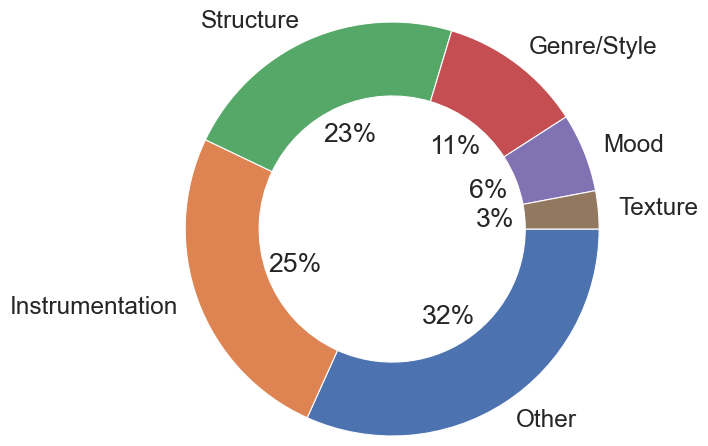}}%
    \vspace{-8mm}
\end{figure}

\paragraph{Data collection} \label{sec:sd_collection}
The captions making up SDD were collected through an open-source online annotation platform \cite{manco_song_2022} aimed at non-experts and volunteers.
Participants were invited via a public announcement circulated on mailing lists and online communities for researchers focused on audio, music, and machine learning, as well as university students more broadly. Most participants were from English-speaking countries (US, GB, IN) and Western Europe (ES, FR, DE, IT), as shown in Figure \ref{fig:countries}. Online participation required participants to be at least 18 years of age, with no additional requirements. We surveyed the participants for their background information for post-hoc quality control and analysis, by collecting answers to questions about their English writing ability (3-point Likert scale), age range, and country of residence. 
To evaluate the familiarity and music knowledge of the participants, we used selected questions from the Goldsmiths Musical Sophistication Index (Gold-MSI) \cite{Mullensiefen2014} with 7-point Likert scale answers to the following statements: (1) \textit{I spend a lot of my free time doing music-related activities [doing]}; (2) \textit{I enjoy writing about music, for example on blogs or social media [writing]}; (3) \textit{I often read or search the internet for things related to music [reading]}.
The analysis of this data together with the code to reproduce the figures is available online.\footnote{\href{https://github.com/mulab-mir/song-describer-dataset}{https://github.com/mulab-mir/song-describer-dataset}}

In the annotation process, annotators were asked to listen to a music recording (up to 2 minutes), write one sentence describing it, and indicate their familiarity with the music on a 3-point Likert scale.
Music recordings were sampled at random from the test split \textit{split-0} of MTG-Jamendo while giving preference to popular tracks as measured by the play count on the Jamendo platform.
A recording could be captioned by up to five participants.
Through this, we gathered 1106 captions from 142 volunteers. The main task instruction, together with an example of multiple captions of the same track provided by different annotators is shown in Table \ref{tab:sd_examples}.
Finally, additional details on the data collection process are given in the datasheet \cite{gebru_datasheets_2021} in Appendix \ref{appendix:datasheet}.

\paragraph{Data quality validation}
\label{subsec:quality_review}
All submitted captions were manually reviewed by the dataset creators to ensure consistency with the annotation guidelines and overall quality, leading to a smaller, curated subset. Reviewers were instructed to rate whether a submitted caption was valid, invalid or borderline, based on whether it adhered to the annotation guidelines provided: \textit{Write one full sentence; use at least 8 words; include things like: genre, mood, instrumentation, ideal listening situation, emotions it evokes, etc.; don't simply provide a sequence of tags; don't use swear words or other offensive language; don't refer to other songs or artists you may know; don't mention personal events or experiences.}
While we observed that, in most cases, captions were high-quality and followed the guidelines, a significant number of cases did not satisfy the single-sentence requirement. Given their high quality,\footnote{An example of such a caption is: \textit{``A nice backbeat punctuates this upbeat, danceable track. The vocals make this sound like taking a walk on a summer afternoon, enjoying the scenery and saying high to the neighbors. Synthesizers modernize the feel and give us a break from the vocals for a bit. A breakdown follows with piano.''}. An example of an invalid caption is instead: \textit{``mood: uplifting, inspiring. vocals: male singer. instruments: guitar, electronic synthesizers''}} these cases were marked as borderline. While all annotations are included in the full dataset, captions which fulfilled all the requirements are marked as part of the valid subset, to enable more reliable evaluation. The validated subset contains rich captions covering a wide range of music aspects.
In an exploratory analysis, we labelled word stems appearing more than 10 times in the data (after removing stop words) with an aspect label. An overview of this analysis is presented in Figure \ref{fig:facets}, where we show the distribution of the most common aspects. These include \textit{Instrumentation} (such as piano, guitar, vocals), \textit{Structure} (e.g. song, accompaniment, chorus), \textit{Genre/style} (e.g. rock, pop, ambient), alongside others such as \textit{Acoustic qualities}, \textit{Era}, \textit{Language}, \textit{Lyrical content}, \textit{Rhythm}, \textit{Tempo}, and \textit{Usage} (denoted by the label \textit{Other} in the plot).

\section{Benchmarking Music-and-Language Models} \label{sec:experiments}
In this section we demonstrate the use of SDD in benchmarking existing music-and-language models. We compare results to those obtained on MusicCaps, currently the most popular dataset in this domain, and highlight the insights that can be drawn from this cross-dataset evaluation.

\paragraph{Tasks and baselines}
\emph{Music captioning} is the task of generating natural language descriptions of a music audio recording and is typically evaluated through a set of automatic metrics such as BLEU (B) \cite{Papineni2002}, METEOR (M) \cite{Lavie2007}, ROUGE (R) \cite{Lin2004} and, more recently, BERT-score (B-S) \cite{zhang_bertscore_2020}. 
As our baselines for this task, we evaluate two encoder-decoder models, MusCaps~\cite{Manco2021} and LP-MusicCaps~\cite{doh_lp-musiccaps_2023}. MusCaps combines a CNN as audio encoder and an LSTM as text decoder, whereas LP-MusicCaps utilizes a combination of a CNN and transformer as an audio encoder, and a transformer as a text decoder. We train both from scratch on the MusicCaps training split, using beam search decoding with a beam size of 4.

\emph{Text-to-music generation} is the task of creating music audio conditioned on a text prompt. Performance on this task can be evaluated along two dimensions: (1) the quality and diversity of the generated audio, using metrics such as the Frechet Audio Distance (FAD) \cite{kilgour_frechet_2019}, Inception Score (IS) \cite{gurumurthy_deligan_2017} and Kullback–Leibler Divergence (KLD), and (2) the relevance between text and audio. Estimating the latter is more elusive and typically addressed via listening studies. Taking inspiration from the MCC metric proposed in \cite{agostinelli_musiclm_2023}, we define a variant of it based on the CLAP model \cite{wu_large-scale_2023}, which we call CLAP Cycle Consistency (CCC). For our evaluation on this task, among the recent but growing pool of text-to-music generative models, we select Riffusion \cite{seth_forsgren_hayk_martiros_riffusion_2022} and AudioLDM \cite{liu_audioldm_2023} based on the availability of pre-trained weights and diversity in their design. 

In \emph{text-to-music retrieval}, the goal is to use text inputs as queries to retrieve corresponding music items. The predominant approach to this is to use multimodal embedding models trained to map different modalities onto a shared space via contrastive learning \cite{Radford2021, wu_audio-text_2023, Manco2022}. Accordingly, for this task, we select two pre-trained state-of-the-art audio-text embedding models: TTMR~\cite{doh_toward_2022}, trained on the Million Song Dataset~\cite{BertinMahieux2011} and CLAP~\cite{wu_large-scale_2023}, trained on Audioset~\cite{Gemmeke2017} and LAION-Audio-630k~\cite{wu_large-scale_2023}.
We evaluate their performance by computing standard information retrieval metrics on the top $k$ retrieved results: Recall at k (R@$k$) and Median Rank (MedR).

\begin{table*}[t]
    \centering
    \footnotesize
\caption{Baseline results on the \textbf{music captioning} task.}
\label{tab:captioning}
\begin{tabular}{p{2cm}cccccccccccc}
\toprule
\multirow{2}{*}{Model} &     \multicolumn{6}{c}{MusicCaps} & \multicolumn{6}{c}{Song Describer} \\
        \cmidrule(rl){2-7}  \cmidrule(rl){8-13}
 & B\textsubscript{1} & B\textsubscript{2} &  B\textsubscript{3}  & M & R & B-S & B\textsubscript{1} & B\textsubscript{2} &  B\textsubscript{3}  & M & R & B-S  \\
\midrule
 MusCaps  & 22.0 & 10.3 & 5.3 & 17.0 & 22.2 & 83.5 & 9.6 & 3.4 & 1.1 & 11.9 & 12.1 & 83.8 \\
LP-MusicCaps  & 28.6  & 13.9 & 7.8 & 20.7 & 19.3 & 87.1 & 12.7  & 4.0 & 1.3 & 16.7 & 11.9 & 86.0 \\
\bottomrule
\end{tabular}
\vspace{-1mm}
\end{table*}

\begin{table*}[t]
\parbox{0.45\linewidth}{
   \centering
    \small
    \caption{Results of the two baselines we evaluate on the  \textbf{text-to-music generation} task.}
    \footnotesize
    \begin{tabular}{llccccc}
    \toprule
        Model  & FAD $\downarrow$  & IS & KLD $\downarrow$ & CCC  \\ 
    \midrule
     \multicolumn{5}{c}{\textsc{MusicCaps}} \\ 
       Riffusion &   11.28 & 1.25 & 1.66  &  0.19 \\
       AudioLDM  & 31.55 & 1.02 & 2.80  & 0.21 \\ 
       \midrule
        \multicolumn{5}{c}{\textsc{Song Describer}} \\ 
        Riffusion &  3.87  &  1.31  &  0.94 &  0.17 \\
        AudioLDM & 20.74 & 1.01 & 2.28 &  0.15 \\
        \bottomrule
    \end{tabular}
    \label{tab:text2music}
    }
\hfill
\parbox{.45\linewidth}{
\centering
    \centering
    \small
    \caption{Results of the two baselines we evaluate on the \textbf{text-to-music retrieval} task.}
    \footnotesize
        \begin{tabular}{lcccc}
    \toprule
        Model & R@1   & R@5  & R@10  & MedR $\downarrow$  \\ 
        \cmidrule(rl){1-5}
         \multicolumn{5}{c}{\textsc{MusicCaps}} \\
       TTMR   &  0.21 & 1.20 & 2.64	 &	638  \\ 
        CLAP &  3.81	& 10.95 & 17.18	& 82.5	\\  
        \midrule
        \multicolumn{5}{c}{\textsc{Song Describer}} \\ 
        TTMR   &   2.95 & 10.19 & 17.43 & 51  \\ 
        CLAP  &   4.42  &	17.02	& 26.01 & 36  \\   
        \bottomrule
    \end{tabular}
    \label{tab:retrieval}
}
\vspace{-6mm}
\end{table*}

\vspace{-3mm}
\paragraph{Results}
For music captioning, in Table \ref{tab:captioning} we observe that, beyond inter-model differences ascribable to the different model architectures and pre-training, both systems display a sizeable gap between in-domain (MC) and out-of-domain (SDD) performance. Although not in itself surprising, this drop in performance underscores the importance of out-of-domain evaluation sets in providing a more realistic assessment of performance in the wild.
From the text-to-music generation results in Table \ref{tab:text2music}, we observe consistent trends between the two datasets on the audio generation metrics: Riffusion tends to produce higher-quality outputs, which may be explained by the fact that AudioLDM is not exclusively trained on music. However, the trend is inverted for semantic relevance. This implies that care should be taken in determining performance superiority based on a single dataset.
Finally, we present the music-text retrieval results in Table \ref{tab:retrieval}. We observe that CLAP consistently outperforms TTMR, and retrieval scores are higher on SSD compared to MusicCaps, which is expected due to its smaller size ($\sim4$x).
Notably, this trend is less prominent for CLAP, and, when adjusting for dataset size, the difference between the two models becomes less stark. Since CLAP is partly trained on AudioSet, this also suggests that evaluating on in-domain distributions may lead to inflated results.

\section{Conclusion} \label{sec:conclusion}
We have presented the Song Describer dataset, a corpus of audio-caption pairs for music-and-language evaluation, and shown how this can be used to perform a more complete assessment of model performance through cross-dataset evaluation. Given the small scale of the dataset, further data collection would be beneficial in strengthening its reliability. Additionally, this may be combined with an extension of the musical cultures represented in both the set of audio recordings and the pool of annotators, which remain limited in the current version.

{
\small
\bibliography{references}
\bibliographystyle{abbrv}
}

\newpage

\appendix

\section{Datasheet for the Song Describer Dataset (SDD)} 
\label{appendix:datasheet}

\subsection{Motivation}
\begin{itemize}
\item \textbf{For what purpose was the dataset created?} \textit{Was there a specific task in mind? Was there a specific gap that needed to be filled? Please provide a description.}

The dataset was created in order to provide the research community with a public dataset for music-and-language tasks containing audio with permissive licenses. The primary use case for the dataset is the evaluation of machine learning models on tasks such as music captioning, text-to-music generation and music-text retrieval.

\item \textbf{Who created the dataset (e.g., which team, research group) and on behalf of which entity (e.g., company, institution, organization)?}

The dataset is a result of a research collaboration undertaken by Universal Music Group International Limited (part of Universal Music Group), Queen Mary University of London, and Music Technology Group (Universitat Pompeu Fabra). 

\item \textbf{Who funded the creation of the dataset?} \textit{If there is an associated grant, please provide the name of the grantor and the grant name and number.}

The dataset creation was funded by:
\begin{itemize}
\item UK Research and Innovation [grant number EP/S022694/1]
\item Universal Music Group
\item The Musical AI project - PID2019-111403GB-I00/AEI/10.13039/501100011033,
funded by the Spanish Ministerio de Ciencia e Innovación
and the Agencia Estatal de Investigación.
\end{itemize}

\item \textbf{Any other comments?}

No.

\end{itemize}

\subsection{Composition} \label{appendix:datasheet_composition}

\begin{itemize}
\item \textbf{What do the instances that comprise the dataset represent (e.g., documents, photos, people, countries)?} \textit{Are there multiple types of instances (e.g., movies, users, and ratings; people and interactions between them; nodes and edges)? Please provide a description.}

The dataset consists of music audio recordings (each recording of a different music track) associated with music captions (up to 5 captions per recording, written by different participants in the annotation platform). 

\item \textbf{How many instances are there in total (of each type, if appropriate)?} 

There are 1106 captions of 706 recordings in total, written by 142 annotators. We also provide a cleaned subset containing manually validated annotations, consisting of 746 captions of 547 audio recordings, written by 114 annotators.

\item \textbf{Does the dataset contain all possible instances or is it a sample (not necessarily random) of instances from a larger set?} \textit{If the dataset is a sample, then what is the larger set? Is the sample representative of the larger set (e.g., geographic coverage)? If so, please describe how this representativeness was validated/verified. If it is not representative of the larger set, please describe why not (e.g., to cover a more diverse range of instances, because instances were withheld or unavailable).}

The audio associated with music captions is a subset of the larger MTG-Jamendo dataset\footnote{\url{https://mtg.github.io/mtg-jamendo-dataset/}} containing over 55k tracks. Due to the annotation effort, we sampled a random subset of tracks from the test set of the \textit{split-0}. The representativeness of our random subset was validated by comparing the distribution of music tags in our subset to that of the full dataset (see Fig. \ref{fig:tag_distribution}).

\begin{figure}[th]
    \centering
    \includegraphics[width=\textwidth]{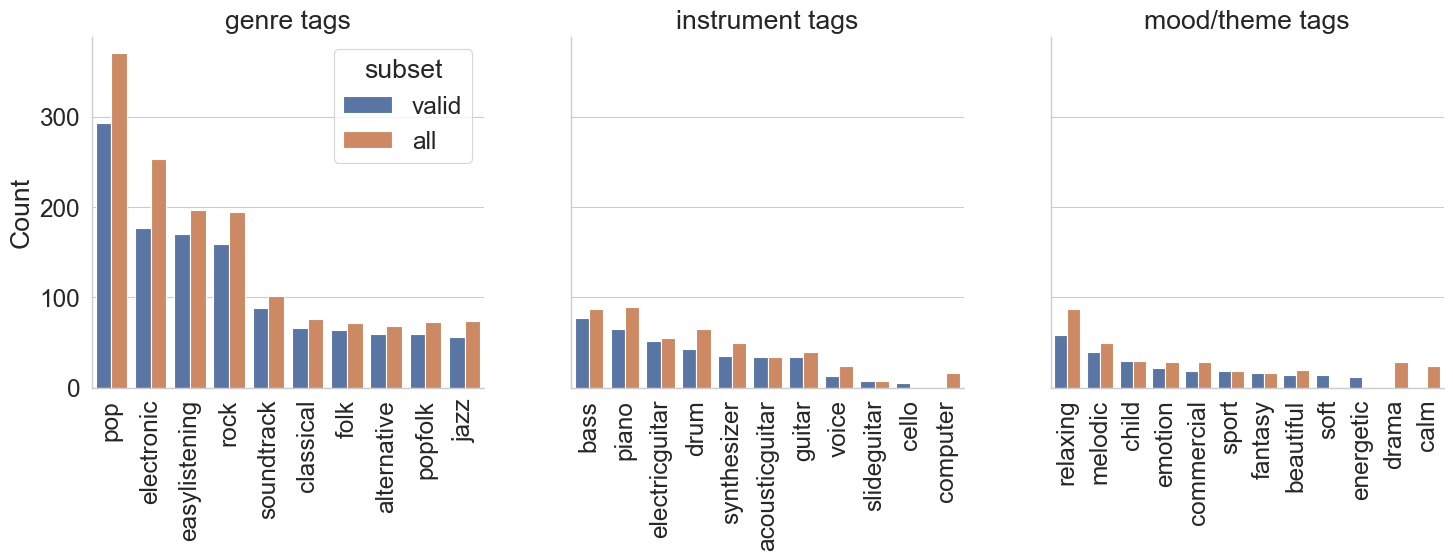}
    \caption{Distribution of the top-10 tags per category taken from MTG-Jamendo in the data collected for SDD.}
    \label{fig:tag_distribution}
\end{figure}

\item \textbf{What data does each instance consist of?} \textit{``Raw'' data (e.g., unprocessed text or images) or features? In either case, please provide a description.}

Each instance consists of an audio-text pair:
\begin{itemize}
\item Audio: one music audio segment per track, with segment duration up to 2 minutes, trimmed from the beginning of a music track (95\% of the segments are 2 minutes). Each audio file is provided in 320kbps 44.1 kHz MP3 audio encoding.
\item Text: unprocessed music caption provided as a list of strings of text, associated with the audio segments.
\end{itemize}

\item \textbf{Is there a label or target associated with each instance?} \textit{If so, please provide a description.}

There is no ground-truth label. Depending on the task, either the audio or the text in a given instance constitutes the target (e.g. in music captioning, the target is the caption).

\item \textbf{Is any information missing from individual instances?} \textit{If so, please provide a description, explaining why this information is missing (e.g., because it was unavailable). This does not include intentionally removed information, but might include, e.g., redacted text.}

There is no missing information.

\item \textbf{Are relationships between individual instances made explicit (e.g., users' movie ratings, social network links)?} \textit{If so, please describe how these relationships are made explicit.}

Each instance is associated to a track ID which can be linked to the original MTG-Jamendo dataset. Relationships between instances can be easily explored from this through the MTG-Jamendo metadata, which contains additional information such as artist, album, genre etc.

\item \textbf{Are there recommended data splits (e.g., training, development/validation, testing)?} \textit{If so, please provide a description of these splits, explaining the rationale behind them.}

There is no recommended split, as the dataset is intended to be used solely for evaluation.

\item \textbf{Are there any errors, sources of noise, or redundancies in the dataset?} \textit{If so, please provide a description.}

The annotations may be prone to noise and bias due to the subjectivity of the task, different levels of familiarity with the kind of music being annotated by the participants, and differences in their English language level. The dataset creators manually reviewed all captions to ensure consistency with the annotation guidelines and overall quality, leading to a smaller, curated subset.

\item \textbf{Is the dataset self-contained, or does it link to or otherwise rely on external resources (e.g., websites, tweets, other datasets)?} \textit{If it links to or relies on external resources, a) are there guarantees that they will exist, and remain constant, over time; b) are there official archival versions of the complete dataset (i.e., including the external resources as they existed at the time the dataset was created); c) are there any restrictions (e.g., licenses, fees) associated with any of the external resources that might apply to a future user? Please provide descriptions of all external resources and any restrictions associated with them, as well as links or other access points, as appropriate.}

The dataset is self-contained (all audio and metadata are distributed as part of the dataset package, including CC licenses for the individual tracks). No fees or restrictions apply to any of the material making up the dataset. The dataset can be additionally mapped to the metadata in the MTG-Jamendo dataset via track IDs. Additionally, the annotation platform used in creating the dataset is open-source and publicly accessible at \url{https://github.com/ilaria-manco/song-describer}.

\item \textbf{Does the dataset contain data that might be considered confidential (e.g., data that is protected by legal privilege or by doctor-patient confidentiality, data that includes the content of individuals' non-public communications)?} \textit{If so, please provide a description.}

There is no confidential information in the dataset.

\item \textbf{Does the dataset contain data that, if viewed directly, might be offensive, insulting, threatening, or might otherwise cause anxiety?} \textit{If so, please describe why.}

Individual songs might contain lyrics that could be considered offensive by listeners. Annotators were appropriately informed about this risk.

\item \textbf{Does the dataset relate to people?} \textit{If not, you may skip the remaining questions in this section.}

The dataset includes human annotations of music recordings and therefore relates to people. However, the participation in the annotation platform was anonymous and no personal data or personally identifiable information was collected. 

\item \textbf{Does the dataset identify any subpopulations (e.g., by age, gender)?}

No.

\item \textbf{Is it possible to identify individuals (i.e., one or more natural persons), either directly or indirectly (i.e., in combination with other data) from the dataset?} \textit{If so, please describe how.}

No.

\item \textbf{Does the dataset contain data that might be considered sensitive in any way (e.g., data that reveals racial or ethnic origins, sexual orientations, religious beliefs, political opinions or union memberships, or locations; financial or health data; biometric or genetic data; forms of government identification, such as social security numbers; criminal history)?} \textit{If so, please provide a description.}

No.

\item \textbf{Any other comments?}

No.
\end{itemize}

\subsection{Collection Process}

\begin{itemize}
\item \textbf{How was the data associated with each instance acquired?} \textit{Was the data directly observable (e.g., raw text, movie ratings), reported by subjects (e.g., survey responses), or indirectly inferred/derived from other data (e.g., part-of-speech tags, model-based guesses for age or language)? If data was reported by subjects or indirectly inferred/derived from other data, was the data validated/verified? If so, please describe how.}

The data was reported by subjects in the form of free-form text annotations and manually validated by the authors following collection.

\item \textbf{What mechanisms or procedures were used to collect the data (e.g., hardware apparatus or sensor, manual human curation, software program, software API)?} \textit{How were these mechanisms or procedures validated?}

The data was collected via a self-developed online software platform.\footnote{\href{https://song-describer.streamlit.app/}{https://song-describer.streamlit.app/}} This was validated via two pilot studies prior to starting the data collection.

\item \textbf{If the dataset is a sample from a larger set, what was the sampling strategy (e.g., deterministic, probabilistic with specific sampling probabilities)?}

The recordings were randomly sampled from the test set of split-0 of the MTG-Jamendo with weights proportional to play counts from the Jamendo platform.\footnote{\href{https://developer.jamendo.com/v3.0}{https://developer.jamendo.com/v3.0}}

\item \textbf{Who was involved in the data collection process (e.g., students, crowdworkers, contractors) and how were they compensated (e.g., how much were crowdworkers paid)?}

Volunteers (above 18 years of age) were recruited via an open call circulated among relevant mailing lists and social networks. We hosted several rounds of annotation with a leaderboard showing (nicknames of) the participants and their number of annotations. The top-3 contributors for each round were rewarded with prizes. These consisted of online vouchers of a value ranging from £10 to £100. 

\item \textbf{Over what timeframe was the data collected? Does this timeframe match the creation timeframe of the data associated with the instances (e.g., recent crawl of old news articles)?} \textit{If not, please describe the timeframe in which the data associated with the instances was created.}

The annotations were collected between 25th November 2022 and 14th April 2023. The instances are associated with music recordings created between 2004 and 2017.

\item \textbf{Were any ethical review processes conducted (e.g., by an institutional review board)?} \textit{If so, please provide a description of these review processes, including the outcomes, as well as a link or other access point to any supporting documentation.}

The project was approved by the Queen Mary Ethics of Research Committee (QMERC reference number: QMERC22.062). As part of the review, the following documentation was provided: a participant information sheet, a consent form, and an application form with detailed questions about the data collection procedure and potential risks. These were assessed by the committee through the low-risk approval route, according to the guidelines outlined on the website.\footnote{\href{http://www.jrmo.org.uk/performing-research/conducting-research-with-human-participants-outside-the-nhs/applications-and-approval/}{http://www.jrmo.org.uk/performing-research/conducting-research-with-human-participants-outside-the-nhs/applications-and-approval/}}

\item \textbf{Does the dataset relate to people?} \textit{If not, you may skip the remaining questions in this section.}

Yes.

\item \textbf{Did you collect the data from the individuals in question directly, or obtain it via third parties or other sources (e.g., websites)?}

We collected the data via the online annotation platform we developed for this purpose (see questions above).

\item \textbf{Were the individuals in question notified about the data collection?} \textit{If so, please describe (or show with screenshots or other information) how notice was provided, and provide a link or other access point to, or otherwise reproduce, the exact language of the notification itself.}

Yes, the platform was built for data collection, thus it includes explicit notice and instructions. The landing page with most of the information can be accessed at \url{https://song-describer.streamlit.app}. A more detailed description of the annotation platform can be found in \cite{manco_song_2022}. Below is the text that informs user about data collection and terms of use.

\begin{quote}
    Song Describer is a tool for crowdsourced collection of music captions, built by a team of researchers from Queen Mary University of London and Universitat Pompeu Fabra Barcelona.

    You will be asked to listen to some music and write a new description for it (Annotation) or evaluate an existing one (Evaluation). Each task takes only a few minutes. You can complete either task as many times as you'd like.
    
    By contributing to Song Describer, you are helping us release the first public dataset of music captions. We really appreciate your help!
    
    We will only collect data you provide by submitting your answers and no other personal information. To find out more about the project and how the data will be used, read our FAQs below.
\end{quote}

\begin{quote}
This survey is part of a research project being undertaken by Universal Music Group International Limited (part of Universal Music Group), Queen Mary University of London, and Universitat Pompeu Fabra (together the “Research Team”).

By participating in this survey, you acknowledge and agree to the following:

• You must be aged 18 or over

• You are participating in an academic study, the results of which may or may not be published in an academic journal.

• Your participation is voluntary and you are free to leave at any time

• All intellectual property rights which may arise or inure to you as a result of your participation in this survey are hereby assigned jointly, in full and in equal proportion to the members of the Research Team.

• By participating in this study, you agree to waive any moral rights of authorship that you may have in the responses that you provide in the survey to the extent permitted by law.

• The data collected by this survey is intended to be published and shall be freely available to all. The responses submitted by you shall not be attributable to you and your participation in the survey shall remain confidential.
\end{quote}

\item \textbf{Did the individuals in question consent to the collection and use of their data?} \textit{If so, please describe (or show with screenshots or other information) how consent was requested and provided, and provide a link or other access point to, or otherwise reproduce, the exact language to which the individuals consented.}

See the link and text in the previous question. At the bottom of the webpage is a button with "Get Started" text that shows a pop-up on hover: ``By clicking here, you confirm that you agree with the statements above.''

\item \textbf{If consent was obtained, were the consenting individuals provided with a mechanism to revoke their consent in the future or for certain uses?} \textit{If so, please provide a description, as well as a link or other access point to the mechanism (if appropriate).}

Participants were made aware of the option to withdraw from the annotation procedure at any point.

\item \textbf{Has an analysis of the potential impact of the dataset and its use on data subjects (e.g., a data protection impact analysis) been conducted?} \textit{If so, please provide a description of this analysis, including the outcomes, as well as a link or other access point to any supporting documentation.}

No analysis of the potential impact was conducted.

\item \textbf{Any other comments?}

No.
\end{itemize}

\subsection{Preprocessing, Cleaning, and/or Labeling}
\begin{itemize}
\item \textbf{Was any preprocessing/cleaning/labeling of the data done (e.g., discretization or bucketing, tokenization, part-of-speech tagging, SIFT feature extraction, removal of instances, processing of missing values)?} \textit{If so, please provide a description. If not, you may skip the remainder of the questions in this section.}

As described in Sec. \ref{subsec:quality_review}, the annotations were manually validated by the research team. This was done by evaluating each caption on a 3-point Likert scale based on its adherence to the annotation guidelines provided. Although we distribute all the annotations (without any preprocessing or cleaning), validated captions are also marked as part of a higher-quality subset.

\item \textbf{Was there data saved in addition to the preprocessed/cleaned/labeled data (e.g., to support unanticipated future uses)?} \textit{If so, please provide a link or other access point to the raw data.}

All data is provided as part of the dataset release.

\item \textbf{Is the software used to preprocess/clean/label the instances available?} \textit{If so, please provide a link or other access point.}

We release code to preprocess, clean and analyse the data in an accompanying GitHub repository.\footnote{\label{repo} \href{https://github.com/mulab-mir/song-describer-dataset/}{https://github.com/mulab-mir/song-describer-dataset/}}

\item \textbf{Any other comments?}

No.
\end{itemize}

\subsection{Uses}
\begin{itemize}
\item \textbf{Has the dataset been used for any tasks already?} \textit{If so, please provide a description.}

The dataset has been used for evaluating machine learning models on the following tasks: music captioning, text-to-music generation, and music-text retrieval.

\item \textbf{Is there a repository that links to any or all papers or systems that use the dataset?} \textit{If so, please provide a link or other access point.}

Links to papers and systems that use the dataset will be added to the GitHub repository.\textsuperscript{\ref{repo}}

\item \textbf{What (other) tasks could the dataset be used for?}

The dataset may additionally be used to study how people describe music. However we note that due to the small sample size and an unbalanced representation of annotators from different countries, the make-up of the annotator pool must be carefully considered before drawing any conclusions from the analysis of the text in the dataset.

\item \textbf{Is there anything about the composition of the dataset or the way it was collected and preprocessed/cleaned/labeled that might impact future uses?} \textit{For example, is there anything that a future user might need to know to avoid uses that could result in unfair treatment of individuals or groups (e.g., stereotyping, quality of service issues) or other undesirable harms (e.g., financial harms, legal risks) If so, please provide a description. Is there anything a future user could do to mitigate these undesirable harms?}

No.

\item \textbf{Are there tasks for which the dataset should not be used?} \textit{If so, please provide a description.}

We aim for this dataset to be used for evaluation and benchmarking of music-and-language models, thus we discourage using it for training.

\item \textbf{Any other comments?}

No.
\end{itemize}
\subsection{Distribution}
\begin{itemize}
\item \textbf{Will the dataset be distributed to third parties outside of the entity (e.g., company, institution, organization) on behalf of which the dataset was created?} \textit{If so, please provide a description.}

The dataset will be publicly available online.

\item \textbf{How will the dataset be distributed (e.g., tarball on website, API, GitHub)?} \textit{Does the dataset have a digital object identifier (DOI)?}

The dataset will be distributed via Zenodo (DOI: \url{https://doi.org/10.5281/zenodo.10072001}).

\item \textbf{When will the dataset be distributed?}

The dataset will be available from 17th November 2023.

\item \textbf{Will the dataset be distributed under a copyright or other intellectual property (IP) license, and/or under applicable terms of use (ToU)?} \textit{If so, please describe this license and/or ToU, and provide a link or other access point to, or otherwise reproduce, any relevant licensing terms or ToU, as well as any fees associated with these restrictions.}

The dataset will be available under the CC BY-SA 4.0 license.\footnote{\url{https://creativecommons.org/licenses/by-sa/4.0/}} The audio files, which are derived from the MTG-Jamendo dataset, are re-distributed as part of the dataset with the respective CC licenses. A list of individual licenses for all audio files is provided with the release.

\item \textbf{Have any third parties imposed IP-based or other restrictions on the data associated with the instances?} \textit{If so, please describe these restrictions, and provide a link or other access point to, or otherwise reproduce, any relevant licensing terms, as well as any fees associated with these restrictions.}

No.

\item \textbf{Do any export controls or other regulatory restrictions apply to the dataset or to individual instances?} \textit{If so, please describe these restrictions, and provide a link or other access point to, or otherwise reproduce, any supporting documentation.}

No.

\item \textbf{Any other comments?}

No.

\end{itemize}

\subsection{Maintenance}
\begin{itemize}
\item \textbf{Who will be supporting/hosting/maintaining the dataset?}

The dataset will be supported and maintained by Queen Mary University of London and the Music Technology Group (Universitat Pompeu Fabra). The dataset will be hosted on Zenodo and supporting code will be hosted on GitHub.

\item \textbf{How can the owner/curator/manager of the dataset be contacted (e.g., email address)?}
Queries about the dataset can be submitted by opening an issue on GitHub or emailing the dataset curators (i.manco@qmul.ac.uk, benno.weck01@estudiant.upf.edu). 

\item \textbf{Is there an erratum?} \textit{If so, please provide a link or other access point.}

An erratum will be provided in the GitHub repository as necessary.

\item \textbf{Will the dataset be updated (e.g., to correct labeling errors, add new instances, delete instances)?} \textit{If so, please describe how often, by whom, and how updates will be communicated to users (e.g., mailing list, GitHub)?}

The dataset will be updated to correct errors if necessary. Additionally, it may be updated with new instances if further annotations are provided by volunteers through our data collection platform. If created, future versions of the dataset will be released via Zenodo and updates will be communicated via both GitHub and Zenodo.

\item \textbf{If the dataset relates to people, are there applicable limits on the retention of the data associated with the instances (e.g., were individuals in question told that their data would be retained for a fixed period of time and then deleted)?} \textit{If so, please describe these limits and explain how they will be enforced.}

No personal data was collected and therefore there are no applicable limits on the retention of the data.

\item \textbf{Will older versions of the dataset continue to be supported/hosted/maintained?} \textit{If so, please describe how. If not, please describe how its obsolescence will be communicated to users.}

Older versions of the dataset will continue to be hosted and maintained. However any code related to the dataset will be updated to support only the most recent version. This will be explicitly mentioned in the repository.

\item \textbf{If others want to extend/augment/build on/contribute to the dataset, is there a mechanism for them to do so?} \textit{If so, please provide a description. Will these contributions be validated/verified? If so, please describe how. If not, why not? Is there a process for communicating/distributing these contributions to other users? If so, please provide a description.}

The website to collect captions is currently available online\footnote{\href{https://song-describer.streamlit.app/}{https://song-describer.streamlit.app/}} and can be accessed for additional contributions. Its source code is also public.\footnote{\href{https://github.com/ilaria-manco/song-describer}{https://github.com/ilaria-manco/song-describer}} Contributions are not directly integrated into the dataset, but will be considered for future version releases.

\item \textbf{Any other comments?}

No.

\end{itemize}

\end{document}